%% file: main_rev2.tex
\documentclass[journal]{IEEEtran} 
\usepackage{graphicx}
\usepackage{subcaption} 
\graphicspath{ {./Figures/} }
\usepackage{multicol, blindtext} 
\usepackage{array}
\usepackage{amssymb} 
\usepackage{amsmath} 
\usepackage[linesnumbered,ruled,vlined]{algorithm2e}
\usepackage{algpseudocode}
\usepackage{amsthm}	
\usepackage[inline]{enumitem} 
\newtheorem{theorem}{Theorem}
\newtheorem{lemma}{Lemma}

\def \EE {\mathbf{E}}
\def \ee {\mathbf{e}}
\def \FF {\mathbf{F}}
\def \GG {\mathbf{G}}
\def \HH {\mathbf{H}}
\def \Heff {\HH_{\text{eff}}}
\def \II {\mathbf{I}}
\def \nn {\mathbf{n}}
\def \PP {\mathbf{P}}

\def \TT {\mathbf{T}}
\def \RR {\mathbf{R}}
\def \UU {\mathbf{U}}
\def \uu {\mathbf{u}}
\def \VV {\mathbf{V}}
\def \tr {\text{tr}}
\def \Ree {\Re \text{e}}
\def \Imm {\Im \text{m}}
\def \yy {\mathbf{y}}
\def \diag {\text{diag}}
\def \rank {\text{rank}}
\def \sig {\sigma^2}

\def \Phii {\mathbf{\Phi}}

\def \Thetaa {\mathbf{\Theta}}
\def \thetaa {\boldsymbol{\theta}}
\def \Lambdaa {\mathbf{\Lambda}}

\def \mh {\hat{m}}
\def \nh {\hat{n}}
\def \Ph {\hat{\PP}}
\def \psih {\hat{\psi}}
\def \psit {\tilde{\psi}}

\input{orcid}

\begin{document}
\title{Low-Complexity Sum-Capacity Maximization for Intelligent Reflecting Surface-Aided MIMO Systems}
\author{
	Ahmad Sirojuddin\orcidA{},
	Dony~Darmawan~Putra\orcidB{} ,~\IEEEmembership{Student Member,~IEEE},
	and Wan-Jen Huang\orcidC{},~\IEEEmembership{Member,~IEEE}
	\thanks{The authors are with the Institute of Communications Engineering, National Sun Yat-Sen University, Kaohsiung 804, Taiwan (email: \{sirojuddin, dony, wanjen.huang\}@g-mail.nsysu.edu.tw)}
}
	
\markboth{IEEE WIRELESS COMMUNICATIONS LETTERS, VOL. xxx, NO. xxx, }%
{Shell \MakeLowercase{\textit{et al.}}: Bare Demo of IEEEtran.cls for IEEE Journals}
\maketitle
	
\begin{abstract}
	Reducing computational complexity is crucial in optimizing the phase shifts of Intelligent Reflecting Surface (IRS) systems since IRS-assisted communication systems are generally deployed with a large number of reflecting elements (REs). This letter proposes a low-complexity algorithm, designated as Dimension-wise Sinusoidal Maximization (DSM), to obtain the optimal IRS phase shifts that maximizes the sum capacity of a MIMO network. The algorithm exploits the fact that the objective function for the optimization problem is sinusoidal w.r.t. the phase shift of each RE. The numerical results show that DSM achieves {\color{black}near maximal sum-rate} and faster convergence speed than two other benchmark methods.
\end{abstract}
\IEEEpeerreviewmaketitle
	
\section{Introduction}
\IEEEPARstart{I}{n} emerging mobile communication systems, various techniques introduces additional degrees of freedom to the transmission systems in order to exploit the multiplexing gain and boost the achievable rate, e.g. large-scale MIMO systems. However, the achievable rate is actually upper bounded by the correlation among the antennas. To introduce higher degrees of freedom, a new technology known as Intelligent Reflecting Surface (IRS) has recently been proposed to shift the phase of the incident signal by a controllable amount \cite{Chongwen19}. IRS consists of a metamaterial integrated with electronic circuits, including a controller, that can alter some of the electronic component properties in such a way as to achieve the desired phase shift amount \cite{Samith20}. IRS thus provides the ability to enhance the throughput of the wireless communication system by adjusting the phase shifts dynamically as required to achieve the constructive interference of the received signal at the destination \cite{LiYou21}.

Numerous studies have considered the use of IRS in improving the throughput of wireless network. For example, Xiu et al. \cite{YueXiu21} maximized the secrecy rate of an IRS-aided millimeter-wave system with low-resolution digital-to-analog converters (DACs). Wang et al. \cite{JunWang21} considered a two-way relay network wherein an IRS was deployed to enhance the minimum capacity of the two users. Guo et al. \cite{Guo20} examined the problem of maximizing the weighted sum-rate in an IRS-assisted multiuser MISO system subject to both perfect and imperfect channel state information (CSI). To reduce  the computational complexity, {\color{black}the authors in \cite{Boyu20} derived a simplified criterion, namely sum-path-gain maximization (SPGM), and utilized} and alternating direction method of multipliers (ADMM) algorithm to {\color{black} enhance} the sum rate of the IRS-aided MIMO systems \cite{Boyu20}.

Although the ADMM method reduced the computational complexity, it came with the price of the degraded sum rate {\color{black}since the ADMM maximizes sum-path-gain indirectly}. With this regard, we propose a novel algorithm designated as Dimension-wise Sinusoidal Maximization (DSM) to solve the {\color{black}SPGM} problem in \cite{Boyu20}, in order to boost the sum rate. For the well-known of an IRS phase-shift of $\phi=e^{j\theta}$, DSM operates directly on the variable of interest $\theta$, which is real, rather than $\phi$, which is complex and restricted by a unity absolute value. Notably, by regarding $\theta$ as the variable of interest, the problem can be solved using the traditional zero-gradient technique. Hence, {\color{black}As demonstrated by the simulation results, the proposed DSM algorithm achieves near maximal sum capacity with faster convergence and lower complexity.}

\section{System Model and Problem Formulation}
\begin{figure}
	\centering
	\includegraphics[width=0.55 \columnwidth]{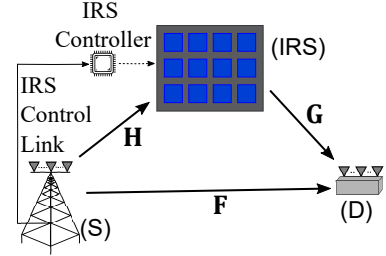}
	\caption{System Model.} 
	\label{fig:SystemModel}
\end{figure}

Consider a MIMO communication system wherein a $K$ antenna-equipped source (S) transmits $U$ parallel data streams, denoted by $\uu \in \mathbb{C}^{U\times 1}$ with $\mathbb{E}\left[\uu\uu^H\right] =\II_U$, $U\leq K$, to an $L$ antenna-equipped destination (D) with the assistance of an IRS consisting of $N$ reflecting elements (see Fig. \ref{fig:SystemModel}). S processes the signal through linear precoding, $\TT\in\mathbb{C}^{K\times U}$, and then broadcasts it to IRS and D simultaneously. Let $\HH\in\mathbb{C}^{N\times K}$, $\GG\in\mathbb{C}^{L\times N}$ and $\FF\in\mathbb{C}^{L\times K}$ be the channels from S to IRS, IRS to D, and S to D, respectively. The IRS shifts the phases of the incident signal by an amount $\thetaa=[\theta_1,\cdots,\theta_N]^T\in\mathbb{R}^N$. The post-processed received signal at D is thus obtained as
\begin{equation} 
	\yy = \left(\sqrt{P/U}\right) \RR\left(\FF+\GG\Phii\HH\right)\TT\uu + \nn,
\end{equation}
where $P\geq 0$ is the total transmitted power, $\Phii=\diag\left[\phi_1, \cdots, \phi_N\right]$ with $\phi_n=e^{j\theta_n}, \forall n\in\mathcal{N} =\left\{1,\cdots,N\right\}$, $\RR\in\mathbb{C}^{U\times L}$ is the linear post-coding matrix at D, and $\nn\sim\mathcal{CN} \left(0,\II_L\right)$ is zero-mean additive Gaussian noise. This letter aims to maximize the end-to-end rate by jointly optimizing the pre-coding $\TT$, post-coding $\RR$, and IRS phase shifts (PSs) $\thetaa$ subject to the available power $P$ at S. The value of $U$ is set to be $U=\rank(\Heff)$ in order to maximally exploit the MIMO spatial diversity, where $\Heff = \FF+\GG\Phii\HH$. The end-to-end capacity from S to D is then given by
\begin{equation} \label{eq:capacity1}
	R = \log_2 \det\left(\II_L + \frac{P}{\sig U} \Heff\TT \TT^H\Heff^H  \right).
\end{equation}

In a MIMO system, the pre-coding matrix $\TT$ and post-coding matrix $\RR$ can be obtained from the singular value decomposition (SVD) of the given $\Heff$. Specifically, let the SVD of $\Heff$ be given by $\Heff=\UU\Lambdaa\VV^H$, where $\UU\in\mathbb{C}^{L\times L}$ and $\VV\in\mathbb{C}^{K\times K}$ are unitary, and $\Lambdaa\in\mathbb{R}^{L\times K}$ is diagonal with positive diagonal entries given by the singular values in descending order, i.e., $\lambda_1 \geq\lambda_2 \geq\cdots \geq\lambda_U$. Hence, the optimal post-coding and pre-coding matrices are given by $\RR^{\text{opt}}=\UU^H$ and $\TT^{\text{opt}}=\VV\Ph^{1/2}$, respectively, where $\Ph=\left[\diag(p_1, \cdots, p_U), \mathbf{0}_{U\times K-U}\right]^T \in\mathbb{R}^{K\times U}$, and $p_u\geq0$ denotes the transmitted power allocation (PA) to the $u$-th data stream, $u\in\mathcal{U}=\left\{1,\cdots,U\right\}$. According to this scheme, the effective channel, $\Heff$, is equivalent to $U$ parallel single-input single-output (SISO) spatial paths, with $\lambda_u$ being the channel of the $u$-th path. The capacity of the $u$-th eigen-channel can be re-expressed as $\hat{R}_u = \log_2\left(1+\frac{Pp_u\lambda_u^2}{\sig U}\right)$. The problem of maximizing the end-to-end rate of the system thus turns into that of maximizing the sum capacity of the equivalent SISO channels. That is,

\begin{align} \label{prob:main}
	\begin{array}{rrclll}
		\displaystyle \max_{\Ph,\thetaa} & \multicolumn{3}{l} {R = \sum_{u=1}^{U} \log_2\left(1+\frac{Pp_u\lambda_u^2}{\sig U}\right)} \\
		\textrm{s.t.} & \UU\Lambdaa\VV^H & = & \FF+\GG\Phii\HH, & \lambda_u = \left[\Lambdaa\right]_{uu}, & \\
		& \left[\Phii\right]_{nn} & = & e^{j\theta_n},  & \forall n\in\mathcal{N}, & \\
		& \sum_{u=1}^{U} p_u& = & U. & &
	\end{array}
\end{align}
The optimal power allocation, $\Ph$, can be obtained by the water-filling (WF) procedure for a fixed $\thetaa$ \cite{Proakis2007}. Specifically, the optimal $p_u$ can be obtained as
\begin{equation}
	p_u = \max\left(\eta - \frac{\sig U}{P\lambda_u^2}, 0\right),
\end{equation}
where $\eta$ is a constant set in such a way as to meet the sum power constraint which can be found numerically.

It can be seen from (\ref{prob:main}) that $\lambda_u, \forall u\in\mathcal{U}$ indicates the quality of the effective channel, $\Heff$, since $\hat{R}_u$ monotonically increases with $\lambda_u, \forall u\in\mathcal{U}$. However, it is difficult to find the optimal value of $\thetaa$ in (\ref{prob:main}) due to its implicit relation with $\lambda_u, \forall u\in\mathcal{U}$ via the SVD of $\Heff$ and the non-linear operation of the power allocation (PA). Thus, to determine the value of $\thetaa$ which maximizes the sum rate of the considered MIMO system, this letter uses the sum-path-gain maximization (SPGM) criterion proposed in \cite{Boyu20} to tackle this implicit relation.

\section{Phase Shifts Design under SPGM Criterion} \label{sec:methods}
To facilitate the optimization problem, this article employs the SPGM criterion, which aims to maximize the sum of the eigen-channel gains, i.e. $\sum_{u=1}^{U} \lambda_u^2$, to enhance the quality of the effective channel $\Heff$. It is noted that this criterion was shown in \cite{Boyu20} to be the lower bound of the sum-rate in (\ref{prob:main}) under an equal PA assumption. Since $\lambda_u^2$ is the $u$-th eigenvalue of $\Heff^H \Heff$, it follows that $\sum_{u=1}^{U} \lambda_u^2 = \tr\left(\Heff^H \Heff\right)$. Accordingly, the optimization of $\thetaa$ can be formulated as
\begin{align} \label{prob:theta}
	\begin{array}{rrclll}
		\displaystyle \max_{\thetaa} & \multicolumn{4}{l} {\psi\left(\thetaa\right) = \tr\left[\left(\FF+\GG\Phii\HH\right)^H \left(\FF+\GG\Phii\HH\right)\right]} \\
		\textrm{s.t.} & \left[\Phii\right]_{nn} & = & e^{j\theta_n},  & \forall n\in\mathcal{N}. &
	\end{array}
\end{align}

The problem in (\ref{prob:theta}) is not convex since the equality constraint is not affine. Thus, this article solves (\ref{prob:theta}) using a self-developed algorithm, designated as the 'Dimension-wise Sinusoidal Maximization' (DSM) algorithm, whose concept is similar to that of the well-known Block Coordinate Descent algorithm \cite{Xu13}. In particular, DSM exploits the fact that, for a particular $n$, $\psi(\thetaa)$ is sinusoidal w.r.t. $\theta_n$ given fixed $\theta_{\nh}, \forall \nh\in\mathcal{N}\backslash n$ (see Lemma \ref{lem:1} below). DSM hence alternately maximizes $\psi(\thetaa)$ w.r.t. each element of $\thetaa$, i.e. $\theta_n, \forall n\in\mathcal{N}$. It is shown later in this article that $\psi(\thetaa)$ is block-sinusoidal w.r.t. $\theta_n$, and the optimal $\theta_n$ can be obtained by a closed-form solution {\color{black}when the other phases are given.}

{\color{black}To facilitate the DSM algorithm, let us start with} a {\color{black}more general} case wherein $\Phii\in \mathbb{C}^{M\times N}$ is a full matrix rather than a diagonal matrix. Under this more stringent assumption, Lemma \ref{lem:1} and Theorem \ref{thm:1} are developed, which can then be applied to more general scenarios. The proposed DSM algorithm is then introduced to solve the optimization problem in (\ref{prob:theta}) by incorporating the Theorem \ref{thm:1} with a diagonal constraint on $\Phii$. It is noted that the underlying equations that construct DSM for both cases of $\Phii$ are quite similar except the element-indices.
\begin{lemma} \label{lem:1}
	Consider the following problem
	\begin{align} \label{prob:thetaLemma}
		\begin{array}{rrclll}
			\displaystyle \max_{\Thetaa} & \multicolumn{4}{l} {\psih\left(\Thetaa\right) = \normalfont{\tr}\left[\left(\FF+\GG\Phii\HH\right)^H \left(\FF+\GG\Phii\HH\right)\right]} \\
			\textrm{s.t.} & \left[\Phii\right]_{mn} & = & e^{j\theta_{mn}},  & \forall m\in\mathcal{M}, n\in\mathcal{N}, &
		\end{array}
	\end{align}
	where $\FF\in\mathbb{C}^{L\times K}$, $\GG\in\mathbb{C}^{L\times M}$, $\Thetaa\in\mathbb{R}^{M\times N}$, $\HH\in\mathbb{C}^{N\times K}$, $\mathcal{M}=\left(1,\cdots, M\right)$ and $\mathcal{N}=\left\{1,\cdots,N\right\}$. The objective function $\psih\left(\Thetaa\right)$ is block-sinusoidal w.r.t. $\theta_{mn}, \forall m\in\mathcal{M}, \forall n\in\mathcal{N}$, with a period of $2\pi$, {\color{black} \textit{i.e.}, $\psit\left(\theta_{mn}\right) = \psih(\Thetaa) \rvert_{\theta_{\mh\nh}}$ is sinusoidal w.r.t. $\theta_{mn}$ given $\theta_{\mh\nh}, \forall \mh\neq m, \forall\nh\neq n$.} 
\end{lemma}
\begin{proof}
	See appendix \ref{apx:Lem1}.
\end{proof}
{\color{black}From the proof in Appendix A, it shows that $\psit(\theta_{mn})$ has two extrema within one block-period, which satisfy the following conditions:
\begin{subequations}
	\begin{align}
		\theta_{mn}^{(\normalfont{\text{opt}})} &= \angle\left[\GG^H\left(\FF+ \GG\tilde{\Phii}_{mn}\HH\right)\HH^H\right]_{mn}, \label{eq:lem1:a}\\
		\theta_{mn}^{(\normalfont{\text{opt}})} &= \angle\left[\GG^H\left(\FF+ \GG\tilde{\Phii}_{mn}\HH\right)\HH^H\right]_{mn} + \pi, \label{eq:lem1:b}
	\end{align}
\end{subequations}
where $\tilde{\Phii}_{mn}\in\mathbb{C}^{M\times N}$ is a matrix with entries of $\Phii$ with the exception of that the $(m, n)$-th entry being zero. Because of the sinusoidal property of $\psit(\theta_{mn})$, one extremum is a maximum and the other is minimum. Theorem 1 further specifies the extrema by exploring their convexity/concavity.
}
\begin{theorem} \label{thm:1}
	{\color{black}Among two extrema described in Lemma 1, the one satisfying (\ref{eq:lem1:a}) achieves global maximum, while the other satisfying (\ref{eq:lem1:b}) achieves global minimum.}
\end{theorem}
\begin{proof}
	See appendix \ref{apx:Thm1}.
\end{proof}

Based on the derived solution of the optimization problem in (\ref{prob:thetaLemma}), the problem in (\ref{prob:theta}) can be regarded as a special case of that in (\ref{prob:thetaLemma}), wherein $M=N$ and $\Thetaa$ is diagonal. Adopting the result in Theorem \ref{thm:1}, the off-diagonal entries in $\Thetaa$ are set as zero and the $n$th diagonal entries, $\theta_n$, are iteratively updated using (\ref{eq:lem1:a}), $n=\{1,\cdots,N\}$. The updating procedure can be viewed as optimizing the individual variables $\theta_n, \forall n\in\mathcal{N}$ which are coupled within $\thetaa$, where the optimal $\theta_n$ in each step is obtained via a closed-form solution. Since $\Thetaa$ is a diagonal matrix, {\color{black}(\ref{eq:lem1:a}) can be re-expressed as
\begin{align} \label{eq:update theta n}
	\theta_n^{(i)} = &\angle \Bigg[  f_{nn} + \sum_{\nh=1}^{n-1} \exp \left(j\theta_{\nh}^{(i)}\right) g_{n\nh}h_{\nh n} \nonumber \\
	&+ \sum_{\nh=n+1}^{N} \exp \left(j\theta_{\nh}^{(i-1)}\right) g_{n\nh}h_{\nh n}\Bigg],
\end{align}
where $f_{nn} \triangleq \left[\GG^H\FF\HH^H\right]_{nn}$, $g_{n\nh} \triangleq \left[\GG^H\GG\right]_{n\nh}$, $h_{\nh n} \triangleq  \left[\HH\HH^H\right]_{\nh n}, \forall n,\nh\in\mathcal{N}$, and $i$ denotes the iteration index. Eq (\ref{eq:update theta n}) has a complexity of $\mathcal{O}\left(N-1\right)$ since it comprises $N-1$ multiplications and $N-1$ additions.
}

{\color{black}Because the condition of $\theta_n$ in (\ref{eq:update theta n}) is comprised of all the remaining angle $\theta_{\nh}$, $\forall \nh\neq n$, we propose Algorithm 1 to approach the optimal solution iteratively.} Algorithm \ref{alg:main} illustrates the basic steps in the method proposed in the present study for maximizing the sum capacity of the system shown in Fig. \ref{fig:SystemModel}. As shown, the algorithm consists of two one-time-executed procedures, namely DSM to obtain the optimal phase-shift, $\thetaa$, (line \ref{alg:main:startDSM} - \ref{alg:main:endDSM}) and water filling (WF) to determine the optimal power allocation ,$\Ph$, and the desired output $R$ (line $\ref{alg:main:WF}$). Lines {\color{black} \ref{alg:main:startfor}-\ref{alg:main:endfor} consists of $N$ times execution of (\ref{eq:update theta n}), thus DSM has a complexity of $\mathcal{O}\left(N\left(N-1\right)\right)$ for each iteration $i$. } WF (line \ref{alg:main:WF}) has a linear complexity, $\mathcal{O}\left(N\right)$, when using the method proposed in \cite{Khakurel14}.

\begin{algorithm}
	\caption{Finding The Global Maximum of Problem \ref{prob:main}}
	\label{alg:main}
	\SetAlgoLined
	\KwIn{$\GG, \HH, \FF, P/\sigma^2$.}
	\KwOut{$R$.}
	Initialize $i=0$. Set an initial $\thetaa$. \label{alg:main:startDSM}\\
	Compute $f_{nn}=\left[\GG^H\FF\HH^H\right]_{nn}$, $g_{n\nh}=\left[\GG^H\GG\right]_{n\nh}$, and $h_{\nh n}=\left[\HH\HH^H\right]_{\nh n}$. \label{alg:main:cons}\\
	\While{$\left|\psi\left(\thetaa^{(i)}\right) - \psi\left(\thetaa^{(i-1)}\right)\right| > \varepsilon$}{
		Increase $i$.\\
		\For{$n=1:N$}{ \label{alg:main:startfor}
			Update $\theta_{n}^{(i)}$ using (\ref{eq:update theta n}). \label{alg:main:eq}\\
		} \label{alg:main:endfor}
	}\label{alg:main:endDSM}
	Compute $\Heff = \FF+\GG\Phii\HH$. $U=\text{rank}(\Heff)$. Get $\UU, \Lambdaa, \VV$ via SVD of $\Heff$. \label{alg:main:Heff} \\
	Obtain $R$ by solving (\ref{prob:main}) with $\Ph$ as the variable of interest using WF procedure. \label{alg:main:WF}
\end{algorithm}

\section{Numerical Results}
This section compares the performance, convergence rate, and complexity of the {\color{black}existing} methods for solving the optimization problem in (\ref{prob:theta}), namely DSM (the proposed method), Gradient Ascent (GA) \cite{Boyd04}, {\color{black}and} Alternating Direction Method of Multipliers (ADMM) \cite{Boyu20}. The steps required to solve (\ref{prob:theta}) using GA are the same as those given in Algorithm \ref{alg:main} with the exception of replacing the statements in lines \ref{alg:main:startfor} to \ref{alg:main:endfor} with $\thetaa^{(i)} = \thetaa^{(i-1)} \pm \alpha \cdot \partial \psi(\thetaa)/\partial \thetaa$, where $\partial \psi(\thetaa)/\partial \thetaa$ is deducted from (\ref{eq:df1/dtheta 2}) as
\begin{equation} \label{eq:dpsi dthetaVect}
	\frac{\partial \psi(\thetaa)}{\partial \thetaa} = -2\, \diag\left(\Imm\left\{\Phii\circ\left[\HH\left(\FF +\GG\Phii\HH\right)^H\GG\right]^T\right\}\right).
\end{equation}
{\color{black}The detailed procedure to solve (\ref{prob:theta}) using ADMM} {\color{black}has been} explained in \cite{Boyu20}. {\color{black}Since SPGM criterion may lead to sub-optimal solution to the maximization of sum capacity in the problem (3), we apply exhaustive grid search that maximizes the sum capacity directly, in order to provide a benchmark of the global optimum. Particularly, the phase of each entry in $\thetaa$ is quantized with step size $2\pi/Q$  in the exhaustive grid search. In the following simulations, the value of $Q$ is set  as $Q=2049$.}

{\color{black}Because it is highly possible that the channel from S to IRS or the channel from IRS to D exists the line-of-sight (LOS) path, the channel matrices $\GG$ and $\HH$ are modeled as Rician fading channel, containing LOS and Non-LOS components. On the other hand, the channel from S to D is assumed Rayleigh fading channel, which simply contains Non-LOS term. In the following simulations, antenna numbers are set as $K=16$ and $L=12$. The LOS component is set as the uniform linear array configuration, while the entries in the Non-LOS components are i.i.d. complex Gaussian distributed with zero mean unit variance. The Rician factors for both $\GG$ and $\HH$ are set as $\beta =10$ dB \cite{Boyu20}}. The other parameters are assigned the specific values shown in the figures.

\begin{figure}
	\centering
	\includegraphics[width=0.75 \columnwidth]{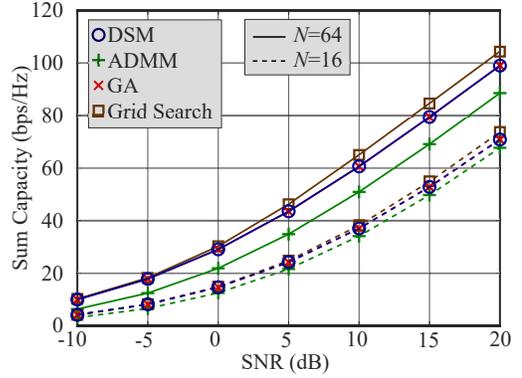}
	\caption{{\color{black}Comparison of spectrum efficiency} given $N=64$ and $N=16$.} 
	\label{fig:SE vs P}
\end{figure}

\begin{figure}
	\centering
	\includegraphics[width=0.75 \columnwidth]{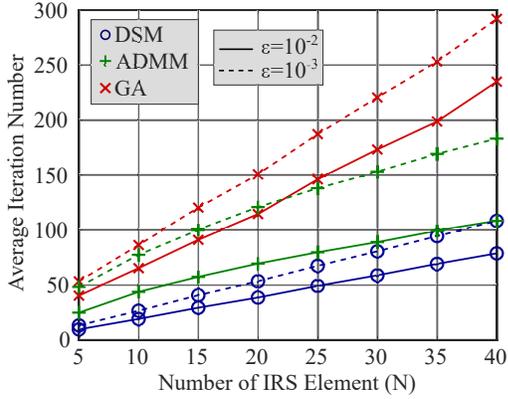}
	\caption{{\color{black}Comparison of the average} iteration {\color{black}numbers} {\color{black}required to achieve} two termination criteria {\color{black} $\varepsilon=10^{-2}$ and $\varepsilon=10^{-3}$}.}
	\label{fig:nIter}
\end{figure}
The performance of the {\color{black}four} methods was evaluated in terms of three metrics, namely {\color{black}the spectral efficiency, the complexity, and the convergence rate}. Figure \ref{fig:SE vs P} demonstrates the spectral efficiency of the {\color{black}four} schemes as a function of SNR for $N=16$ and $N=64$. It shows that the DSM and GA methods achieve comparable spectral efficiency and both outperform the ADMM method at all SNR and $N$. The performance gap between them increases with $N$. {\color{black} It shows that the proposed DSM and GA method achieve tight performance gap compared with the exhaustive grid search.}

{\color{black} {\color{black}Because the procedures of the DSM and GA are similar, the complexities of the} DSM and GA {\color{black}are of the same order, equal to} $\mathcal{O} \left( (L+K)N^2 +KL(L+N) +I_D N(N-1)\right)$ and $\mathcal{O} \left( (L+K)N^2 +KL(L+N) +I_G N^2\right)$, respectively, where $I_D$ and $I_G$ are the iteration {\color{black}numbers} for DSM and GA, respectively. On the other hand, {\color{black}the} ADMM has a complexity of $\mathcal{O} \left(KL(L+N^2) +N^3 +I_A N^3 \right)$, where $I_A$ is the iteration number for the {\color{black} ADMM \cite{Boyu20}. In general, the DSM, ADMM and GA methods demand polynomial complexity in terms of $K$, $L$ and $N$. It shows that} per-iteration complexity of {\color{black}the} DSM is {\color{black}slightly} lower than that of GA and {\color{black}much lower than that of } ADMM. The {\color{black} exhaustive grid search} has a complexity of $\mathcal{O} \left( Q^N KL(L+N)\right)$.}

To evaluate the convergence rate, Figure \ref{fig:nIter} shows the average iteration {\color{black}numbers} of the {\color{black}three} considered methods {\color{black}i.e., the values of $I_D$, $I_G$, and $I_A$} as a function of $N$ for two different termination criteria: {\color{black}$\varepsilon=10^{-2}$ and $\varepsilon=10^{-3}$}. Note that to ensure fairness, $\FF$, $\GG$, $\HH$ and the initial values of $\thetaa$ are assigned the same values in all three schemes for each channel realization. In the GA method, the step-size $\alpha$ used in Fig. \ref{fig:nIter} is chosen from a pre-defined set to minimize the required iteration number for each value of $N$ \cite{Boyd04}. The result shows that DSM converges with {\color{black}the fewest iterations} for both termination criteria. Furthermore, ADMM requires fewer iterations than GA. The slope of GA is steeper than that of either DSM or ADMM, which indicates that GA is unsuitable for systems with large $N$. Notably, GA has step-size issue and thus $\alpha$ must be properly assigned in order to avoid slow convergence or even divergence. 

\begin{figure}
	\centering
	\includegraphics[width=0.75 \columnwidth]{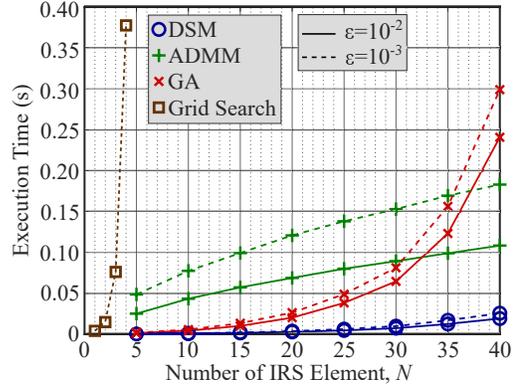}
	\caption{\color{black} Comparison of the execution time to achieve two termination criteria.}
	\label{fig:time}
\end{figure}

{\color{black}To corroborate the low complexity of the proposed method, Figure \ref{fig:time} {\color{black} compares} the execution time of {\color{black} the} DSM {\color{black}with} others given a channel realization. {\color{black}Although the DSM and GA methods have similar per-iteration complexity, the proposed DSM converges faster and requires much  lower number of iterations. As a result, the DSM demands the shortest execution time, while the complexity of the GA significantly increases with $N$. The} ADMM consumes stably increasing time with $N$, but its execution time is still longer than that of the DSM. The {\color{black}exhaustive} grid search undoubtedly demands {\color{black}prohibitively} high computation time {\color{black}since its complexity grows exponentially with $N$}.}

In summary, the results show that DSM outperforms GA and ADMM in solving (\ref{prob:theta}) in terms not only of an improved solution performance, but also a faster convergence speed and a lower complexity.

\section{Conclusion}
This letter has proposed a novel low-complexity Dimension-wise Sinusoidal Maximization (DSM) algorithm for determining the optimal IRS phase shifts which maximizes the sum capacity of a single-point MIMO system. The proposed algorithm follows the same principal as the well-known Block Coordinate Descent (BCD) method in decoupling the variable of interest and then optimizing each variable in turn. The numerical results have shown that compared to the ADMM method proposed in \cite{Boyu20} and the general Gradient Ascent method, DSM not only achieves a better solution for the IRS phase shift optimization problem, but also has a faster convergence speed and a lower complexity in each iteration.

\appendices
\section{Proof of Lemma \ref{lem:1}} \label{apx:Lem1}
{\color{black} To prove the function $\psih (\theta_{mn})$ is sinusoidal, let us first explore the first derivative given by
\begin{align}
    \frac{\partial \psih\left(\Thetaa\right)}{\partial \theta_{mn}} =& \tr\left[\HH^H \frac{\partial \Phii^H}{\partial \theta_{mn}}\GG^H\left(\FF +\GG\Phii\HH\right)\right] \nonumber \\
	& +\tr\left[\left(\FF +\GG\Phii\HH\right)^H\GG\frac{\partial \Phii}{\partial \theta_{mn}}\HH\right] \nonumber \\
	=& 2\Ree\left\{\tr\left[je^{j\theta_{mn}}\left(\FF +\GG\Phii\HH\right)^H \GG \ee_m\ee_n^T \HH\right]\right\} \label{eq:df1/dtheta 1} \\
	=& -2\Imm\left\{e^{j\theta_{mn}}\left[\HH\left(\FF +\GG\Phii\HH\right)^H\GG\right]_{nm}\right\} \label{eq:df1/dtheta 2}
\end{align}
Eq. (\ref{eq:df1/dtheta 2}) exploits the fact that the trace of matrix multiplication is invariant under cyclic permutations and $\Ree \left\{jc_1\right\} =-\Imm\left\{c_1\right\}$ for any complex $c_1$. The matrix $\Phii$ can be further decomposed as $\Phii =\tilde{\Phii}_{mn} + e^{j\theta_{mn}}\EE_{mn}$, where $\EE_{mn} = \ee_m \ee_n^T \in \mathbb{R}^{M\times N}$, $\ee_m$ is a basis vector with the $m$th entry being one and the others being zero, and $\tilde{\Phii}_{mn}$ is an $M\times N$ matrix with entries of $\Phii$ with the exception of the $(m, n)$-th entry being zero. Hence, the first derivative of $\psih(\theta_{mn})$ can be written by
\begin{align}
    \frac{\partial \psih\left(\Thetaa\right)}{\partial \theta_{mn}} =&-2\Imm\Biggl\{e^{j\theta_{mn}}\Big[\HH\left(\FF +\GG\tilde{\Phii}_{mn}\HH\right)^H \GG \nonumber\\
	&+e^{-j\theta_{mn}}\HH\left(\GG\EE_{mn}\HH\right)^H\GG\Big]_{nm}\Biggr\} \nonumber \\
	= & -2\Imm\left\{e^{j\theta_{mn}} \left[\GG^T\left(\FF +\GG \tilde{\Phii}_{mn}\HH\right)^*\HH^T\right]_{mn}\right\} \label{eq:df1/dtheta 3}
\end{align}
Eq. (\ref{eq:df1/dtheta 3}) exploits the fact that $\left[\HH\left(\GG\EE_{mn} \HH\right)^H\GG\right]_{nm} =\ee_n^T \HH\left(\GG\ee_m\ee_n^T\HH \right)^H\GG\ee_m =\left[ \HH\HH^H\right]_{nn} \left[ \GG^H\GG\right]_{mm} \triangleq r_{mn}$ is real, and $\Imm\left\{c_2+r_{mn}\right\} = \Imm\left\{c_2\right\}$ for any complex $c_2$. Denote $z_{mn} = \left[\GG^T\left(\FF +\GG \tilde{\Phii}_{mn}\HH\right)^*\HH^T\right]_{mn}$, then (\ref{eq:df1/dtheta 3}) can be simplified as
\begin{equation} \label{eq:df1/dtheta 4}
    \frac{\partial \psih\left(\Thetaa\right)}{\partial \theta_{mn}} = -2\Imm\left\{e^{j\theta_{mn}} z_{mn}\right\}.
\end{equation}
Note that $z_{mn}$ depends on all angles except $\theta_{mn}$. Hence, given all angles except $\theta_{mn}$, the first derivative of $\partial \psih / \partial \theta_{mn}$ is sinusoidal w.r.t. $\theta_{mn}$ with a period of $2\pi$. Each block period contains two extrema points, which are the global maximum and minimum due to the nature of the sinusoidal function. In addition, the extrema points can simply be found from the fact that the solutions to $\Imm\left\{e^{j\theta_{mn}} z_{mn}\right\}=0$ are $\theta_{mn} = \angle z_{mn}^* +2\pi k$ or $\theta_{mn} = \angle z_{mn}^* +\pi +2\pi k$ where $k$ is an integer.
}

\section{Proof of Theorem \ref{thm:1}} \label{apx:Thm1}
Taking the second derivative of $\psih\left(\Thetaa\right)$ w.r.t. $\theta_{mn}$ (by taking the first derivative of (\ref{eq:df1/dtheta 1})) yields
{\color{black}
	\begin{align}
		\frac{\partial^2 \psih\left(\Thetaa\right)}{\partial \theta_{mn}^2} =& 2\Ree\Bigl\{-e^{j\theta_{mn}}\ee_n^T\HH\left(\FF +\GG\Phii\HH\right)^H\GG\ee_m \nonumber \\
		& +je^{j\theta_{mn}}\ee_n^T\HH\HH^H(-je^{-j\theta_{mn}})\EE_{mn}^T \GG^H\GG\ee_m\Bigr\} \nonumber \\
		=& 2\Ree\Bigl\{-e^{j\theta_{mn}}\left[\HH\left(\FF +\GG\Phii\HH\right)^H\GG\right]_{nm} \nonumber \\
		& +\left[\HH\HH^H\right]_{nn}\left[\GG^H\GG\right]_{mm}\Bigr\} \nonumber \\ 
		=& -2\Ree \Bigl\{ e^{j\theta_{mn}} \left[\HH\left(\FF +\GG\tilde{\Phii}_{mn} \HH\right)^H\GG\right]_{nm} \Bigr\} \nonumber \\
		& -2\Ree \Bigl\{ e^{j\theta_{mn}} \left[\HH\left( \GG e^{j\theta_{mn}} \EE_{mn} \HH\right)^H\GG\right]_{nm} \Bigr\} \nonumber \\
		&+ 2\Ree \Bigl\{ \left[\HH\HH^H\right]_{nn}\left[\GG^H\GG\right]_{mm} \Bigr\} \label{eq:d2psi dtheta2 before cancel} \\
		=& -2\Ree \left\{ e^{j\theta_{mn}} z_{mn}\right\}, \label{eq:d2psi dtheta2 final}
	\end{align}
	where (\ref{eq:d2psi dtheta2 final}) holds since the second and third terms of (\ref{eq:d2psi dtheta2 before cancel}) cancel each other.
	
	To get the extremum maximum, $\partial^2 \psih\left(\Thetaa\right) / \partial \theta_{mn}^2$ must be negative since it is located in the concave region. Hence, the requirement to obtain $\theta_{mn}$ that maximize $\psih(\Thetaa)$ is expressed as
	\begin{equation} \label{eq:real zero 1}
	    \Ree\left(e^{j\theta_{mn}} z_{mn}\right) > 0,
	\end{equation}
	with {\color{black}an opposite sign} to obtain the extremum minimum.
	}
	
	To determine which expression ((\ref{eq:lem1:a}) or (\ref{eq:lem1:b})) leads to the global maximum or global minimum, $\theta_{mn}$ given by (\ref{eq:lem1:a}) or (\ref{eq:lem1:b}) can be substituted into (\ref{eq:real zero 1}) and a check then made of the satisfied inequality. For example, suppose that $\theta_{mn}$ is given by (\ref{eq:lem1:a}), {\color{black}i.e., $\theta_{mn} =\angle z_{mn}^*$}, then (\ref{eq:real zero 1}) yields
	\begin{equation}
		\Ree\left\{e^{j\theta_{mn}} z_{mn}\right\} =\Ree\left\{e^{j\theta_{mn}}|z_{mn}|e^{-j\theta_{mn}}\right\} = |z_{mn}|>0.
	\end{equation}
	In other words, (\ref{eq:lem1:a}) is the condition to obtain the global maximum. Substituting $\theta_{mn}$ obtained from (\ref{eq:lem1:b}), {\color{black}i.e., $\theta_{mn} =\angle z_{mn}^* +\pi$}, into (\ref{eq:real zero 1}) yields
	\begin{align}
		&\Ree\left\{e^{j\theta_{mn}} z_{mn}\right\} 	=\Ree\left\{e^{j\theta_{mn}}|z_{mn}|e^{-j\left(\theta_{mn} -\pi\right)}\right\} \nonumber \\
		&=-|z_{mn}| <0.
	\end{align}
	Hence, (\ref{eq:lem1:b}) is the condition required to obtain the global minimum.

\bibliographystyle{IEEEtran}\vspace{-0.25em}
\bibliography{IEEEabrv,bibliography_ex}

\end{document}

%% file: orcid.tex
\usepackage{tikz,xcolor,hyperref}
\definecolor{lime}{HTML}{A6CE39}
\DeclareRobustCommand{\orcidicon}{%
	\begin{tikzpicture}
	\draw[lime, fill=lime] (0,0) 
	circle [radius=0.16] 
	node[white] {{\fontfamily{qag}\selectfont \tiny ID}};
	\draw[white, fill=white] (-0.0625,0.095) 
	circle [radius=0.007];
	\end{tikzpicture}
	\hspace{-2mm}
}

\foreach \x in {A, ..., Z}{%
	\expandafter\xdef\csname orcid\x\endcsname{\noexpand\href{https://orcid.org/\csname orcidauthor\x\endcsname}{\noexpand\orcidicon}}
}